\documentclass[acmsmall]{acmart}

\acmJournal{PACMHCI}
\acmYear{2022} \acmVolume{6} \acmNumber{CSCW1}
\acmArticle{131} \acmMonth{January} \acmPrice{15.00}
\acmDOI{10.1145/3512978}

\AtBeginDocument{%
  \providecommand\BibTeX{{%
    \normalfont B\kern-0.5em{\scshape i\kern-0.25em b}\kern-0.8em\TeX}}}

\begin{document}

\setcopyright{acmcopyright}
\acmJournal{PACMHCI}
\acmYear{2022} \acmVolume{6} \acmNumber{CSCW1} \acmArticle{131} \acmMonth{4} \acmPrice{15.00}\acmDOI{10.1145/3512978}

\title{\revision{Understanding the Role of Context in Creating Enjoyable Co-Located Interactions}}

\author{Szu-Yu (Cyn) Liu}
\authornote{Research conducted while affiliated with Snap Research.}
\email{cynliu@microsoft.com}
\affiliation{%
  \institution{Microsoft}
  \city{Redmond}
  \state{Washington}
  \country{United States}
}
\affiliation{%
  \institution{University of California, Irvine}
  \city{Irvine}
  \state{California}
  \country{United States}
}

\author{Brian A. Smith}
\email{bsmith@snap.com}
\affiliation{%
  \institution{Snap Inc.}
  \city{Santa Monica}
  \state{California}
  \country{United States}
}
\affiliation{%
  \institution{Columbia University}
  \city{New York}
  \state{New York}
  \country{United States}
}

\author{Rajan Vaish}
\authornote{Co-Principal Investigators.}
\email{rvaish@snap.com}
\affiliation{%
  \institution{Snap Inc.}
  \city{Santa Monica}
  \state{California}
  \country{United States}
}

\author{Andr\'es Monroy-Hern\'andez}
\authornotemark[2]
\email{amh@snap.com}
\affiliation{%
  \institution{Snap Inc.}
  \city{Seattle}
  \state{Washington}
  \country{United States}
}
\affiliation{%
  \institution{Princeton University}
  \city{Princeton}
  \state{New Jersey}
  \country{United States}
}

\renewcommand{\shortauthors}{Szu-Yu (Cyn) Liu et al.}

\def \revision #1{{\textcolor{black}{#1}}}

\begin{abstract}

In recent years, public discourse has blamed digital technologies for making people feel ``alone together,'' distracting us from \revision{engaging with} one another, even when we are interacting in-person. We argue that \revision{in order} to design technologies that foster and augment co-located interactions, we need to first understand the context in which enjoyable co-located socialization takes place. We \revision{address} this gap by surveying and interviewing over 1,000 U.S.-based participants to understand what, where, with whom, how, and why people enjoy spending time in-person. Our findings suggest that people enjoy engaging in everyday activities with \revision{individuals with whom they have} strong social ties because it helps enable nonverbal cues, facilitate spontaneity, support authenticity, encourage undivided attention, and leverage the physicality of their bodies and the environment. We conclude by providing a set of recommendations for designers interested in creating co-located technologies that encourage social engagement and relationship building.

\end{abstract}

\ccsdesc[500]{Human-centered computing~Empirical studies in HCI}

\keywords{co-located interaction; co-located technology; in-person; enjoyment; social}

\maketitle

\section{Introduction}
Information and communication technologies are deeply embedded in the fabric of our day-to-day lives and have drastically changed the way people interact, not only remotely, but also in-person. Previous research has shown that while digital technologies make it easier to stay connected online, they can also disrupt and alienate people in face-to-face social interactions \cite{oduor2016frustrations, turkle2017alone, misra2016iphone}--- to the extent that even when people are physically together, or co-located, they are still ``alone'' in their own ``digital bubbles''  \cite{turkle2017alone, rogers2014bursting}. 

An open challenge for Human-Computer Interaction (HCI), then, is to understand how to design technology that supports rather than inhibits in-person interactions---in other words, \revision{to study} how technology can make co-located casual social interactions more enjoyable and engaging \cite{lindley2005designing, fischer2016collocated}. Following Olsson et al., who recently argued that researchers should focus on designing technologies that increase the quality of co-located social interactions \cite{olsson2019technologies}, we note that, in order to develop such technology, we  must first establish a holistic understanding regarding the \textit{context} in which enjoyable in-person interactions take place.

In this work, we \revision{performed} a mixed-methods study designed to establish a holistic understanding of enjoyable casual in-person interactions, aimed at guiding future design efforts in this area. Our survey design and interview protocols were based on prior work on \textit{context} and focused on the following attributes: relationship \textit{(who)}, activity \textit{(what)}, location \textit{(where)}, time \textit{(when)}, technology \textit{(how)}, and value \textit{(why)} people enjoy interacting with others in-person \cite{Abowd1999Towards}. We deployed this U.S.-only survey with 1,007 participants and performed 27 semi-structured \revision{follow-up} interviews. Our study focused on pre-pandemic scenarios; \revision{this created a number of limitations that we will discuss later} (\S6).

Our findings show how different contextual attributes play a role in making co-located interactions enjoyable. \revision{First,} in terms of relationships, we observe that people most enjoy spending time with their ``strong ties'' \cite{granovetter1977strength}, which include not only \textit{romantic partners} and \textit{friends}, but also \textit{pets}. \revision{In terms of} activities, we find that although \textit{chatting or having conversations} is the favorite activity \revision{of} most people, it is often supported by complementary activities such as \textit{eating}, \textit{drinking}, and \textit{playing games}. \revision{With respect to} location, we note that many of people's preferred in-person activities are \textit{location-independent} (that is, the interaction is not restricted to a specific location), and that \textit{location-dependent} activities often take place in \textit{private} spaces. More importantly, we observe that it is the atmosphere (rather than the location itself) that influences people's preference \revision{of} location types. Technology-wise, we find that devices are often used to initiate activities and then fade into the background; whether it is desirable to include technology \revision{during an interaction} depends on an implicit consensus among the co-located individuals. Finally, our findings suggest that one of the most valued aspects of being physically co-located is that physicality and nonverbal cues help support authenticity, connectedness, and attentiveness; and that it is precisely these elements that make in-person interactions more genuine, intimate, engaging, and \revision{therefore} preferable \revision{to} interacting online.

The contributions of this work are two-fold. \revision{First, through a mixed-method study, we establish} an empirical foundation that illustrates the role of context \revision{in} the in-person interactions people prefer. \revision{Second}, we propose design strategies to support future ideation \revision{surrounding} enjoyable co-located technologies.

\section{Related Work}

As digital devices proliferate in our everyday lives, there is increasing interest in exploring how technology might enhance engagement and enjoyment in casual social scenarios \revision{\cite{alavi2012ambient, balestrini2016jokebox, fischer2016collocated, paasovaara2017understanding, jarusriboonchai2016design, kyto2017augmenting, lucero2012mobicomics}}. In this paper, we follow this research trend and focus on supporting engagement in in-person interactions (\S2.1) and understanding the sociotechnical context in which co-located technology should be designed (\S2.2).

\subsection{Supporting Engagement and Enjoyment in Co-Located Scenarios}

Previous work has shown that technology use in co-located scenarios can undermine conversation quality, increase social conflicts, and reduce relationship satisfaction \cite{przybylski2013can, roberts2016my, wang2017partner, rotondi2017connecting, oduor2016frustrations}. Turkle \cite{turkle2017alone} discusses how heavy dependency on mobile devices \revision{diverts} our attention from the people we interact with and even \revision{replaces} face-to-face encounters with online interactions. In her words, ``We defend connectivity as a way to be close,'' but in reality, it ``also disrupts our attachments to things that have always sustained us,'' including ``genuine'' and ``authentic'' social interactions \cite{turkle2017alone}. Similarly, Pickersgill's photographs \cite{Pickersgillremoved} poignantly illustrate how disruptive technology can be for relationship bonding by simply editing away people's digital devices, showing, for example, a couple in bed ignoring each other while paying attention to their empty hands.

Roberts and David coined the term ``phubbing'' to illustrate the negative encounter of being ``snubbed by someone using their cellphone when in your company'' \cite{roberts2016my}. Studies show that \revision{being phubbed} can lead to a sense of disconnection as an ongoing in-person interaction is considered less important than an incoming notification on a mobile device \cite{ling2008new, przybylski2013can, su2015third}. For example, Wang et al.'s study with 243 married couples suggests that partner phubbing has a negative effect on relationship satisfaction and can increase the risk of depression \cite{wang2017partner}. Similarly, Oduor and colleagues conducted diary studies and found that ``family members become frustrated when others do non-urgent activities on their phones in the presence of others'' \cite{oduor2016frustrations}. Further, Przybylski and Weinstein found that the mere presence of mobile phones can hinder relationship formation 
and undermine closeness and connection between co-located individuals \cite{przybylski2013can}. Overall, we see a need for HCI researchers to address the sociotechnical ``alone together'' problem \cite{turkle2017alone}.

\subsection{Understanding the Sociotechnical Context of Enjoyable Co-located Interactions}

The HCI community has identified different strategies to mitigate technology-induced social alienation, including 
developing technologies that curtail excessive device use in co-located scenarios \cite{lochtefeld2013appdetox, kim2019lockntype, ko2015nugu, okeke2018good} and creating systems that make in-person interactions more enjoyable \cite{fischer2016collocated, gencc2020designing, park2017don, isbister2016yamove, lucero2012mobicomics}. In this paper, our focus is \revision{on} the latter \revision{approach}. 

Prior work on technology-enhanced co-located social interactions has explored a wide range of topics, such as collective play \cite{segura2017design, isbister2016yamove, balestrini2016jokebox, paasovaara2017understanding}, storytelling \cite{van2009collocated, benford2000designing}, photo-sharing \cite{clawson2008mobiphos, lucero2013mobile, nielsen2014juxtapinch}, collaborative virtual reality (VR) and mixed reality (MR) experiences \cite{gugenheimer2017sharevr, yang2018sharespace, zhou2019astaire}, social engagement \revision{\cite{dagan2019designing, memarovic2012using, powell2012table, laureyssens2014zwerm, lucero2011pass, porcheron2016using}}, community building \cite{mccarthy2004augmenting, wozniak2015rufus}, leisure activities \revision{ \cite{durrant2011automics, paasovaara2017understanding, mauriello2014social}}, and self-expression \cite{epp2018augmented, kao2015mugshots, kyto2017augmenting}. 
However, Lundgren et al. noted that prior studies often ``[exist] as isolated exemplars of technical systems'' \cite{lundgren2015designing}. To our knowledge, there is no formative study that provides a deep understanding of the context that enjoyable co-located technology should be designed for.

We \revision{address} this gap and aim to establish a holistic understanding \revision{of} the types of in-person activities people enjoy in order to support future work \revision{in this area.}

Our focus is on enjoyment because\revision{, although} it is a central component for engagement and joy, \revision{it} remains an under-explored \revision{aspect of} co-located technology \cite{lindley2005designing,sweetser2005gameflow}. In this study, we leave the term ``enjoyment'' undefined and open \revision{to} our participants' interpretation to understand whether there are certain features across different in-person interactions that lead to enjoyment.
 
Following Blythe and Hassenzahl \cite{blythe2003semantics}, who note that enjoyment ``doesn't exist in and of itself'' but rather \revision{embodies a} ``context specific'' \revision{quality}, we turn our focus to the \textit{context} of co-located social intercourse to study what makes in-person interactions joyful, engaging, and preferable. We understand context broadly as ``any information that can be used to characterize the situation of an entity. An entity is \revision{defined as} a person, place, or object that is considered relevant to the interaction between a user and an application, including the user and applications themselves'' \cite{Abowd1999Towards}. In this paper, we draw on the four primary attributes of context proposed by Abowd et al. \cite{Abowd1999Towards}---identity \textit{(who)}, activity \textit{(what)}, location \textit{(where)}, and time \textit{(when)}---to guide our investigation on how different contextual dimensions might affect the quality of co-located interactions.

\section{Methods}
Drawing from Abowd et al.'s  primary attributes of context \cite{Abowd1999Towards}, we aim to gain an empirical understanding into how various contextual attributes may affect the quality of in-person social interactions. Our study involved a two-stage inquiry---surveys (\S3.1) and semi-structured interviews (\S3.2). The surveys were conducted first to offer quantitative insights into identifying the in-person interactions people enjoy, including \textit{who} they prefer spending time with, \textit{what} activities they like to do together, \textit{where} they enjoy hanging out, and \textit{whether/what} technologies are involved in their favorite in-person interactions. Next, we conducted semi-structured interviews to gain qualitative insights into questions such as \textit{how} technology use may influence co-located interactions, and \textit{why} people prefer interacting in-person in certain scenarios. Both studies were conducted online to broaden the geographical diversity of our participants.

\subsection{Surveys}
We started with a formative survey to explore the landscape of enjoyable co-located interactions before designing and deploying a large-scale survey aimed \revision{at capturing and ranking the} in-person activities people prefer.

\subsubsection{Formative Survey}
In the first phase, our goal was to identify a comprehensive list of relationship types and activities people enjoy doing when they are physically together. To do so, we deployed an online survey focused on two questions: ``who do you most enjoy spending time in person with ?'' and ``what do you do with them when you spend time together in person?'' The first question was a single-choice question we \revision{adopted} from literature \cite{jones2010feasibility}, but \revision{it offered participants} the option of providing a free response;
the second question was open-ended. To gather a wide range of responses from participants with diverse demographics, we recruited respondents from two distinct online platforms: Amazon Mechanical Turk, where our participants skewed older and male \cite{difallah2018demographics}, and Snapchat, where our participants skewed younger and female \cite{Snapchatuser, Snapchatusergender}. We ran the survey between February 21 and March 2, 2020, and collected 3,434 valid responses. We developed a codebook to analyze the open-ended answers. We then hired three independent coders from Upwork\footnote{Upwork is an online freelancing platform: \url{https://www.upwork.com/}}, who performed open-coding individually to label the responses from a random set of participants. All authors met regularly to discuss the codes until we reached agreement and resolved all coding conflicts. 

\begin{table*}[]
\small
\caption{Self-reported demographics of 1,007 survey participants included in our analysis. Participant demographics resemble\revision{d that} of the U.S. population. \revision{We adopted the age ranges used in by Pew research Center \cite{Pew} to define generational cohorts.}} 
\label{tab:surveydemo}
\centering
\resizebox{\textwidth}{!}{%
\begin{tabular}{ll|ll|ll|ll}
\toprule
\multicolumn{2}{c}{\textbf{Gender}} & \multicolumn{2}{c}{\textbf{Age Bracket}} & 
\multicolumn{2}{c}{\textbf{Race}} & \multicolumn{2}{c}{\textbf{Education}} \\     
\midrule
Female & 50.04\% & Gen Z (13-23) & 17.68\% & American Indian & 0.40\% & $<$ High school & 6.45\%\\
Male & 49.45\% & Millennial (24-39) & 28.60\% & Asian & 5.46\% & High school & 20.85\%\\
Non-binary & 0.20\% & Gen X (40-55) & 24.33\% & Black & 12.02\% & Some college & 25.52\%\\
Other & 0.20\% & Boomer (56-74) & 26.61\% & Hispanic/Latino & 17.68\% & Associates & 14.00\%\\
No answer & 0.10\% & Silent ($\geq$75) & 2.78\% & Two or more races & 3.97\% & Bachelor's & 22.14\%\\
&&&& Pacific Islander & 0.10\% & Graduates & 10.33\% \\
&&&& White & 59.88\% & No answer & 0.70\%\\
&&&& No answer & 0.50\% && \\ 
\midrule
\multicolumn{8}{c}{\textbf{Occupation}} \\
\midrule
Administration & 4.87\% & Engineering & 0.79\% &  Sales or retail & 1.29\% & Grad student & 2.18\%\\
Arts or journalism & 2.18\% & Food services & 1.29\% & Self-employed & 6.85\% & Retired & 18.37\%\\
Business & 9.04\% & Homemaker & 6.06\% & Skilled labor & 5.66\% & Unemployed & 6.45\%\\
CS or IT & 3.48\% & Legal & 0.99\% & Grade school student & 4.77\% & Other & 8.54\%\\
Education & 5.36\% & Medical & 4.87\% & College student & 5.46\% & No answer & 1.49\%\\
\bottomrule
\end{tabular}
}
\end{table*}

\subsubsection{\revision{Main} Survey}

Based on Abowd et al.'s \cite{Abowd1999Towards} definition of context, our survey has five categories: identity \textit{(who)}, activity \textit{(what)}, location \textit{(where)}, technology \textit{(how)}, and time \textit{(when)}. We configured the survey such that \revision{participant responses to} later questions \revision{were dependent} on their answers to prior questions; the survey is included in the Appendix. 

\begin{itemize}
    \item \textbf{Identity:} \revision{Information gathered in this survey section included (1) participant's} self-reported identity (i.e., age, gender, race, education, and occupation), and (2) relationship \revision{type} they have with who\revision{m}ever they most enjoy spending time with in-person. The list entailed 11 relationtypes we ientified in our formative survey.

    \item \textbf{Activity:} \revision{From a list of the top 26 in-person activities drawn from our formative survey, participants} selected and ranked the activities they most enjoy doing with the people they mentioned in the Identity part of the survey.

    \item \textbf{Location:} Participants \revision{were asked if their favorite activity needed to take place in} a specific location or \revision{if it could be done} anywhere (i.e., whether the activity was location-dependent or location-independent), and if the former \revision{was indicated}, what type of place was preferred (for example, public or private).

    \item \textbf{Technology:} \revision{Participants noted} what types of technologies, if any, they use when doing their favorite activity.
    \item \textbf{Time:} Here, participants were asked to describe in detail the last time they hung out in-person with their favorite person. The concept of time organically \revision{arose} from their free-text descriptions of the frequency (e.g., \textit{``we had a weekly gathering''}) and duration (e.g., \textit{``we spent an afternoon''}) \revision{of their favorite shared activity} and the time of the day \revision{during which} the activity took place (e.g., \textit{``we went out to dinner''}).
\end{itemize}

We recruited participants aged 13 years or older who reside in the United States. The survey ran from March 13 to 18, 2020. We used Qualtrics\footnote{Qualtrics: \url{https://www.qualtrics.com/}} for hosting our survey, as well as recruiting and compensating the participants; Qualtrics charged us \$5 USD for each completed survey response we received. The first author worked with Qualtrics to gather a participant pool that resembled the demographics of the United States \cite{ACS} (Table~\ref{tab:surveydemo}). We screened out responses that included 
incomplete or gibberish answers and those whe\revision{re} the respondents reported that they did not enjoy spending time with anyone in-person. In the end, we collected a total of 1,007 valid responses. The survey ended with an invitation to participate in a follow-up interview, which we describe next.

\subsection{Interviews}
We conducted interviews to follow up on the survey results and gather qualitative insights 
that were otherwise \revision{difficult} to obtain from survey \revision{alone}. In total, we conducted 27 interviews between March 18 and April 23, 2020.

\subsubsection{Interview Structure}
We developed an semi-structured interview protocol (see Appendix B) to understand the lived experiences of individuals and, in particular, \revision{to explore the} role context plays in making co-located interactions enjoyable. We began the interviews by collecting the participants' demographic information, including their age and location. Following Abowd et al.'s \cite{Abowd1999Towards} argument that their four contextual factors \revision{(identity, activity, location, and time)} should offer ``indices into other sources of contextual information,'' we started the interviews by following a script but \revision{also used} unscripted questions \revision{to} gather more information about the specific context and experiences of in-person activities mentioned by the participants \cite{hesse2017practice}.

\subsubsection{Participant Interviews}
We recruited 27 people to interview: 16 signed up through the survey and 11 through a recruitment message \revision{distributed via} the authors' institutional and social networks. Each interview lasted 50-70 minutes and was conducted via Google Hangouts videochat (or via phone when slow connectivity was an issue). Participants received a \$20 gift card at the end of the interview. We recorded \revision{and transcribed} the interviews, then we performed a thematic analysis \revision{of} both the transcripts and the notes taken during the interviews \cite{braun2006using}. The first author grouped similar codes into themes. All authors met frequently to refine codes, themes, and concepts.

All interviewees self-reported their age and geographic location. Their ages ranged from 13 to 83; \revision{participants were categorized according to the following demographic cohorts:} 40.74\% Gen Z (aged 13-23 in year 2020), 29.63\% Millennial (aged 24-39), 14.81\% Gen X (aged 40-55), 11.11\% Boomer (aged 56-74), and 3.7\% Silent (aged 75 or older) \cite{Pew}. Our interviewees were located in 15 different states and 5 different time zones in the United States. Only \revision{16 of our interviewees answering the survey indicated their} gender and race: half were females and half were males. Their races were diverse: 12.5\% Asian, 12.5\% Black, 12.5\% Hispanic, 50\% White, 6.25\% mixed races, and 6.25\% other race.

\section{Results}
We organize our results around the five contextual attributes: relationships \textit{(who)}, activities \textit{(what)}, locations \textit{(where)}, technologies\textit{(what/how)}, and value \textit{(why)}. We include findings about time \textit{(when)} \revision{alongside} the other attributes because it is dependent on them, e.g., people enjoy playing games with friends on the weekends. We combine our findings from the surveys (\S3.1.2) and the interviews (\S3.2) to provide both quantitative and qualitative insights of people's preferences. We note that when we report percentages, we base them on our \textit{survey} respondents; when we report quotes, they are from the interviews or free-text \revision{responses given} in the surveys. We refer to our interviewees by identifier P\#.

\begin{figure}[]
  \centering
  \includegraphics[width=\linewidth]{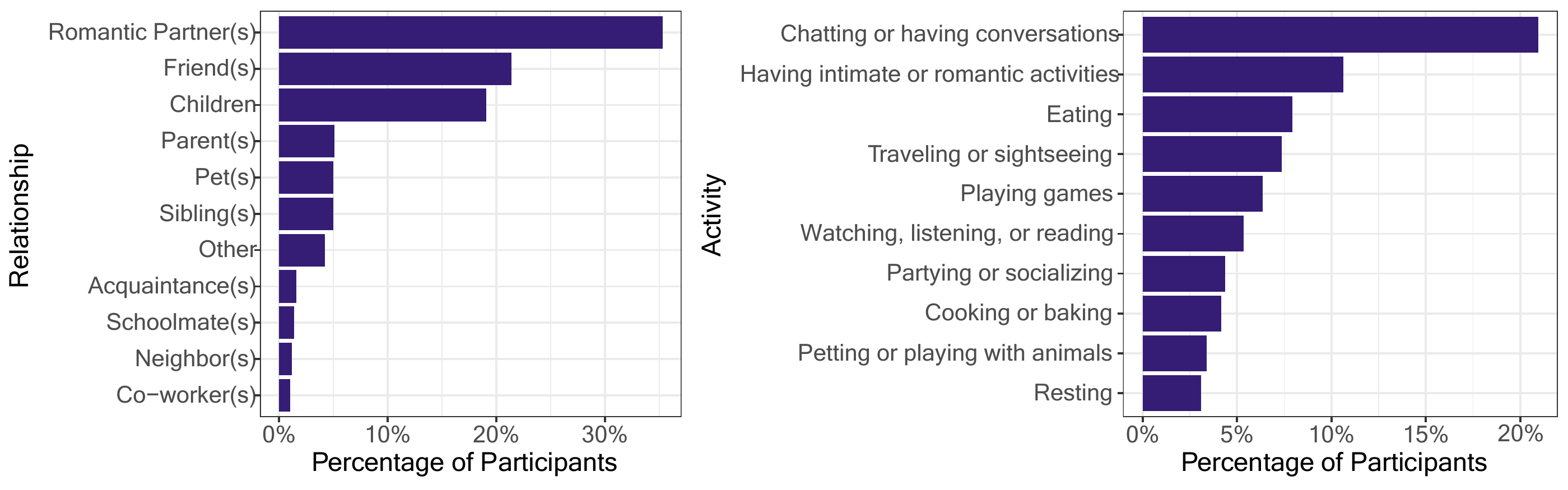}
  \caption{\label{fig:who-what-font} (Left) Survey participants' favorite person to spend time in-person with. (Right) Top ten in-person activities.}
\end{figure}

\subsection{Relationships: strong preference not only for romantic partners, friends, children, but also for pets}
Participants noted that \textit{``people are [the] number one key''} (P26) for making co-located interactions enjoyable. Not surprisingly, our results show that people much prefer hanging out in-person with \revision{individuals they regard as} ``strong ties'' over \revision{those they consider} ``weak ties'' \cite{granovetter1977strength}. The three most preferred relationships for co-located interaction were \textit{romantic partners} (35.3\%), \textit{friends} (21.35\%), and \textit{children} (19.07\%). Interestingly, \revision{the option} \textit{pets} was not originally included as a choice but \revision{arose} organically from the free responses in our formative survey. In the end, \textit{pets} were ranked in the top five on this list, with 4.97\% of our participants reporting \textit{pets} as their favorite individuals to spend time with. Conversely, very few participants reported enjoying hanging out with their ``weak ties'' \cite{granovetter1977strength}, such as acquaintances and co-workers (1.6\% and 1.0\%, respectively).  (Figure~\ref{fig:who-what-font}, Left).

\subsubsection{Children and pets for females; romantic partners for males}
Some results varied by gender and age. For example, more than twice \revision{as many} females \revision{as} males reported \revision{that they} most enjoy spending time with their \textit{children} (25.6\% and 12.65\%, respectively) and \textit{pets} (7.14\% and 2.81\%, respectively). Conversely, significantly more males than females reported enjoying hanging out with \textit{romantic partners} (42.17\% and 30.16\%, respectively). Age also plays a role in participants' preferences for individuals they most enjoy spending time with. For instance, although participants across all age groups prefer hanging out with their \textit{romantic partners}, Gen Zs are an exception: they prefer \textit{friends} (41.57\%) over \textit{romantic partners} (23.03\%). Our Gen Z interviewees \revision{revealed} that they and their friends have \textit{``synchronized''} thoughts and actions (P16) and that they can be \textit{``really goofy''} (P25) without having to worry about being judged.

\subsubsection{Big groups for fun; small groups for intimacy}
Results from our interviews reveal that the number of people participating in an activity also shapes the social atmosphere. Specifically, participants mentioned that activities such as \textit{``go[ing] to the beach''} (P21), \textit{``football games''} (P21), or \textit{``party''} (P17) \revision{are} more \textit{``fun''} and \textit{``eventful''} (P21) with bigger groups. Conversely, our interviewees described small group gatherings as more \textit{``meaningful''} (P24) and \textit{``intimate''} (P27). Our interviewees' preference for group size depends on both the nature of the activity and their personal mood, noting that \textit{``If I'm in a bad mood, I probably don't want to be dealing with a big group interaction... I'd rather have more meaningful interactions with people that I'm closer with''} (P24).

\subsection{Activities: chatting is the central experience, but \revision{it} is often supported by complementary activities}
Figure~\ref{fig:who-what-font} (right) lists the ten most-favored in-person activities\revision{; these} include but are not limited to \textit{chatting or having conversation} (20.95\%), \textit{intimate or romantic activities} (10.63\%), \textit{eating} (7.94\%), \textit{traveling or sightseeing} (7.35\%), and \textit{playing games} (6.36\%). Overall, we see that most \revision{of the} in-person activities participants enjoy are mundane, everyday interactions. Specifically, \textit{chatting or having conversations} was considered the cornerstone \revision{of} relationship building. For instance, P1 told us that  \textit{``most of''} her relationship with her romantic partner \textit{``is built on these, like, spontaneous conversations.''} Interestingly, while \textit{chatting} is the top activity for many, our interview data shows that people rarely chat without also doing something else at the same time. For example, P8 mentioned that he often \textit{eats}, \textit{drinks}, or \textit{plays games} while \textit{chatting} with others to avoid \revision{getting} into intense or awkward conversations:

\begin{quote}
     \textit{
     ``There's [always] some kind of distraction... You might play a video game in the background or you might have some drinks in the background... it might be the main thing that's happening, but it's never the only thing. I think having some kind of other activity eases the awkwardness of it.''}
\end{quote}

\subsubsection{Friends enjoy partying and playing games while couples favor intimate activities and traveling} 
We found that people's favorite in-person activities were highly dependent on whom they spend time with. \revision{For example, our survey shows} that while spending time with friends, participants most enjoy \textit{chatting} (25.58\%), \textit{partying or socializing} (11.63\%), and \textit{playing games} (7.91\%). In comparison, when hanging out with romantic partners, participants reported preferring \textit{intimate or romantic activities} (27.04\%), \textit{chatting} (15.77\%), and \textit{traveling or sightseeing} (10.70\%). What is also interesting is that romantic partners were considered the best people to spend time with while chatting (26.54\%), having intimate activities (89.72\%), eating (30.00\%), and traveling (51.35\%)\revision{; the exception was} playing games, where people prefer playing with their children (32.8\%), friends (26.56\%), siblings (10.94\%), and even pets (10.94\%) over romantic partners (7.81\%).

\subsubsection{Activity initiates in-person encounters, but people come to the foreground afterwards} 
Many interviewees mentioned that the activity itself is what makes co-located interactions possible. For instance, P8 described going to the movies and going for a run as the main \textit{``medium''} to \textit{``bring other people.''} Similarly, P27 noted that they would always suggest an in-person activity whenever they wanted to chat with friends: \textit{``For me, it's kind of weird to just ask somebody, hey, you want to just talk somewhere?''} Intriguingly, we observed that the initiating activity often fades into the background afterwards, to the extent that P23 described \revision{the experience as follows:} \textit{``I don't think it matters too much what we're doing... I think just hanging out with a person, like, no matter what you're doing, is nice.''}

\subsubsection{In-person interaction is preferable but not always an option} 
Our interviewees also mentioned that being able to hang out in-person is a privilege. For example, P5 mentioned that he is \textit{``restricted in the [kind] of activities''} he can \revision{take part in} due to a chronic health condition; others reported that \textit{``there's not much to do''} (P2), \textit{``not much time''} (P24), or \revision{that it's }\textit{``too costly''} to do activities outside (P27). A few also mentioned that having friends and families in \textit{``different places''} makes it difficult to gather in-person. Additionally, weather can also be a limiting factor: \textit{``we don't interact that much in the wintertime, because it's just not practical to sit there and traipse around in a foot of snow with dogs''} (P6). In many cases, we see that face-to-face interaction has indeed become ``a luxury good'' \cite{humancontact}.

\subsubsection{One-of-a-kind activities create memories; everyday interactions build relationships} 
Finally, we observed \revision{that} how often an activity takes place also affects how individuals perceive their co-located experiences. Overall, events that happened less often, such as school trips to Europe, were considered memorable because \textit{``
we are sharing that first-time experience together''} (P26). \revision{Conversely,} everyday activities are often the core to relationship building. For instance, P3 described the daily routines he developed with his romantic partner \revision{as offering} a \textit{``feeling of closeness'':}

\begin{quote}
     \textit{``I was never used to listening to NPR in the morning... [but] it became like routine. She'll play every morning, no matter what... We didn't just build memories, we built routines.''}
\end{quote}

\subsection{Location: space types affect not only the activities, but also how people act and feel}
\begin{figure}[]
  \centering
  \includegraphics[width=\linewidth]{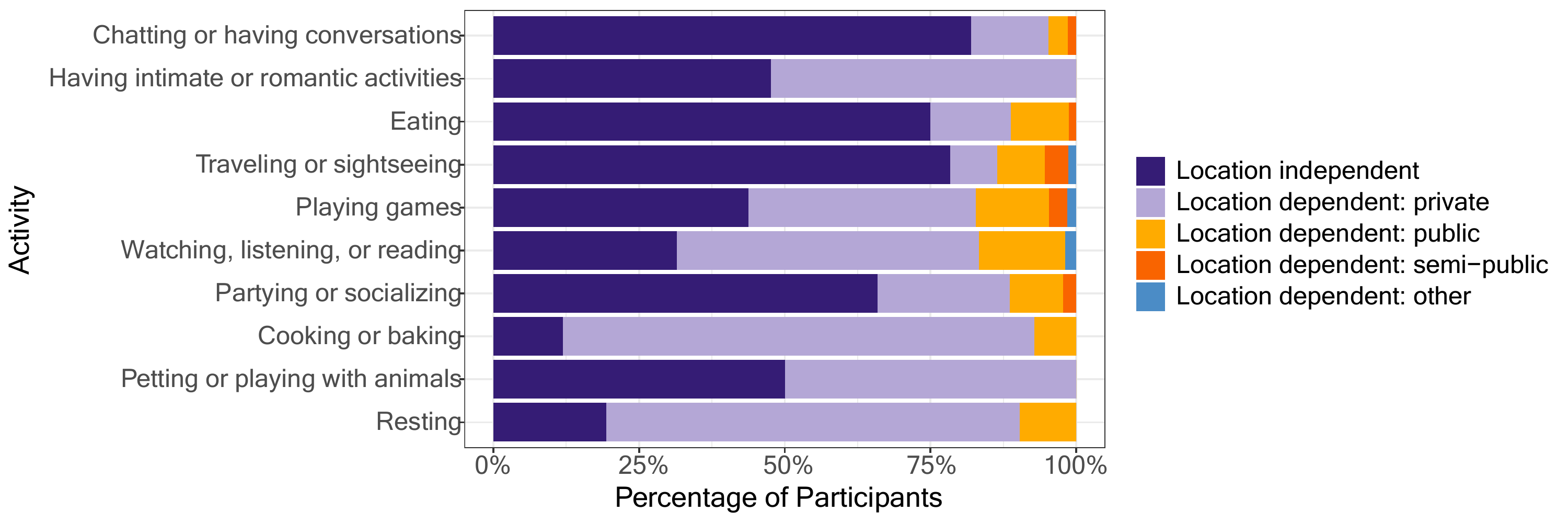}
  \caption{\label{fig:location-font}	Type of location people prefer for each of the most liked in-person activities.}
\end{figure}

We wanted to understand if the activities people enjoy with their favorite person were tied to a specific physical space (location-dependent) or if they could take place anywhere (location-independent). Figure~\ref{fig:location-font} shows the locations our survey participants preferred across different activities. 

\subsubsection{If an activity is not tied to a specific private space, it can happen anywhere}
Participants reported preferring specific locations \revision{in} five of the ten most favored in-person activities. Additionally, for activities identified \revision{as being location-dependent} by the majority of participants, private spaces (e.g., homes or private vehicles) significantly outweighed public spaces (e.g., parks or caf{\'es}). Specifically, people prefer private spaces over public spaces while engaging in \textit{intimate or romantic activities} (52.34\% versus zero percent), \textit{playing games} (39.06\% versus 12.50\%), \textit{watching, listening, or reading} (51.85\% versus 14.81\%), \textit{cooking or baking} (80.95\% versus 7.14\%), and \textit{resting} (70.97\% versus 9.68\%). On the contrary, our survey reveals that location-independent activities are the ones that do not require specific settings or specialized devices to unfold; these activities include chatting (81.99\% participants \revision{described this as a} location-independent activity), eating (75.00\%), traveling (78.38\%), and partying (65.91\%).

\subsubsection{Private spaces support authenticity; public spaces facilitate spontaneity}
Participants \revision{observed} that the space \revision{an activity occurs} affects how they feel and act. In general, interviewees reported feeling comfortable doing whatever they wanted without having to worry about being judged in private spaces, especially at home. For instance, P10 mentioned that \textit{``when we were indoors... we could just act [like] a fool without people looking at us, and when we’re in public, we [have] to be cautious [about] how loud we are and [d]on’t draw attention to [ourselves].}'' Analogously, P8 noted that \textit{``it seems odd... to have a very intense discussion in public}”; instead, he prefers \textit{``go[ing] to somebody’s house''} to engage in a discussion. Similarly, P14 particularly enjoys ``chilling in the den'' with his wife every night after dinner \revision{as they} watch and discuss videos together. This particular setting is critical and gives him \textit{``a feeling of fulfillment and satisfaction''}:

\begin{quote}
     \textit{``If we were sitting in two folding chairs in an empty room, it would feel very cold and sterile and we probably wouldn't get as much out of it. It's weird, atmosphere means a lot. When we're in our den, you know, it's nice and comfortable, we have the dog hanging out with us on the couch... we're in our home and it's comfortable and we're just relaxing.''} --P14
\end{quote}

In contrast, public spaces help facilitate spontaneous encounters. For example, P5 mentioned that the local post office is a great place to have impromptu social interactions because \textit{``the town is so small that there \revision{[are] }no postal workers... [so] you have to go to the post office [to pick up mail].''} Similarly, P6 described \revision{that} a particular bar in the small town he lives in always has live music: \textit{``It draws people there... on Wednesday nights to go and play Queen of Hearts.''}

\subsubsection{Good spaces are defined by the atmosphere and not \revision{the} location alone}
Our findings suggest that it is not the location \textit{per se} that determines the quality of in-person interactions. Instead, location matters because it sets the atmosphere \revision{for} different types of social encounters \revision{to take place} (i.e., it serves as an ``affordance'' \cite{norman1999affordance}). For example, P24 noted that she does not feel comfortable having deep conversations \textit{``when there's blaring music playing''}; she continued, \textit{``If I'm, like, in my room late at night, we would probably have a deeper interaction and a more meaningful interaction.''} In her narrative, both the location type and the time of day \revision{during which} the interaction takes place play a deciding role in establishing the mood. 

Other factors that affect the quality in-person interactions include sound, light, people, activity, and environmental distractions. For some, one is more likely to have deep conversations when the \textit{``noise level''} is low (P24), when there is \textit{``quieter music''} playing (P17), or when \textit{``people are somewhat intoxicated''} (P17). For others, instead of staying at one specific place, \revision{establishing the right atmosphere might} involve \textit{``going for a walk''} to avoid \textit{``distractions''} in a public space, such as a caf{\'e} (P13). In comparison, for those who enjoy playing VR games with others, the\revision{ir preference was for} \textit{``big space[s]''} to avoid bumping into each other (P22), or \textit{``dark''} spaces, \textit{``so it makes it even more fun.''} (P18).

\subsection{Technologies: devices often initiate activities, then fade into the background}
\begin{figure}[]
  \centering
  \includegraphics[width=\linewidth]{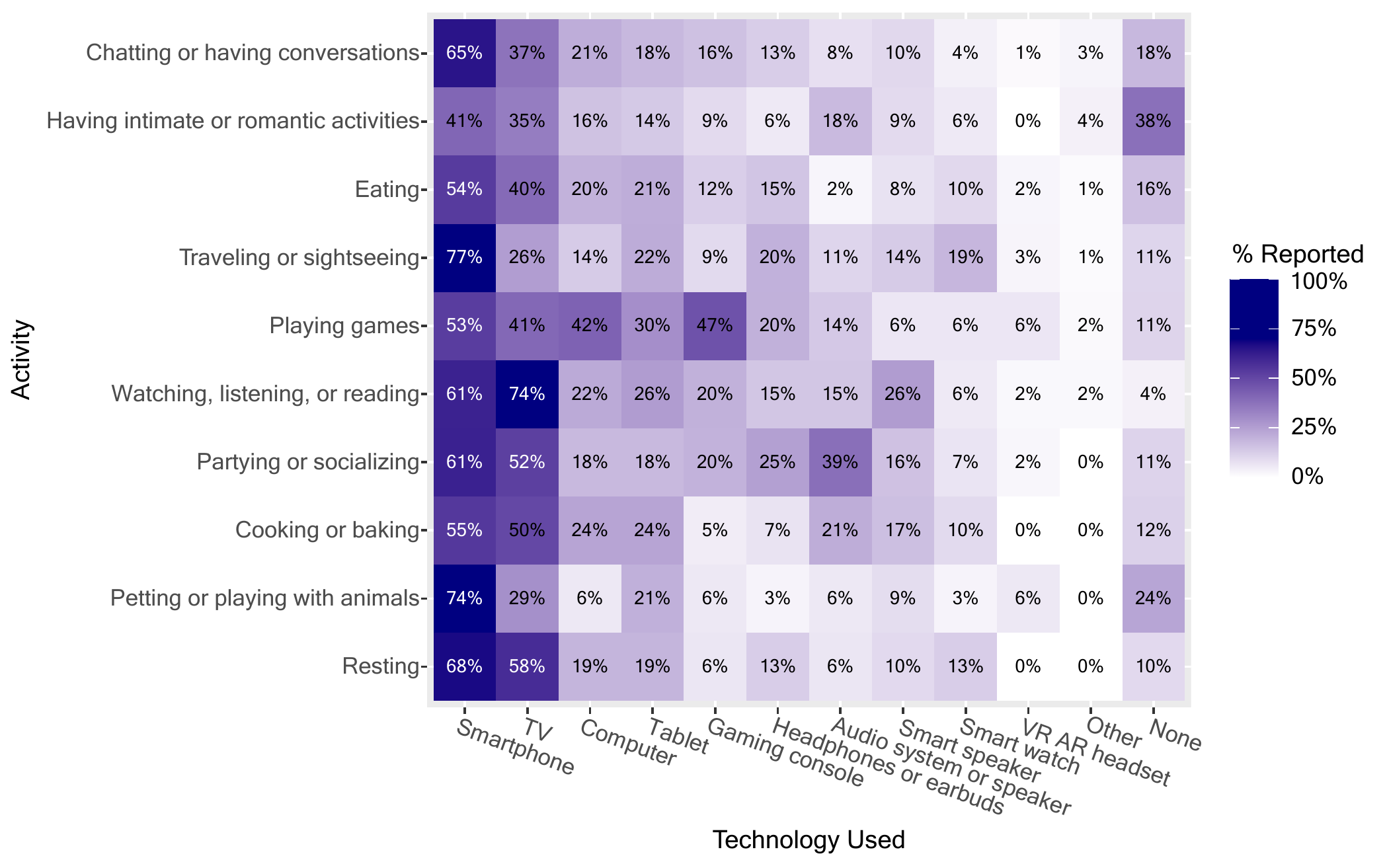}
  \caption{\label{fig:technology-font} Technology use in \revision{the} top ten in-person activities. Smartphones are the most used devices across popular activities. Percentages are rounded to the nearest whole integer.}
\end{figure}

We asked our participants whether there was any technology involved in their favorite co-located activities. We found that the most used devices included \textit{smartphones} (60.87\%), \textit{TVs} (39.13\%), \textit{computers} (21.65\%), and \textit{tablets} (20.46\%). Additionally, \revision{approximately} one in six participants (17.18\%) reported not using any device while engaging in their favorite activities. Figure~\ref{fig:technology-font} shows the popularity of different digital devices \revision{in} the top in-person activities. 

\subsubsection{Males and younger participants mention higher reliance on technologies}
We observe that in co-located interactions, male participants used almost all \revision{types} of digital devices more than females. The gender difference in technology use is especially notable in \textit{audio systems or speakers}; only 9.52\% of females reported using audio systems, the percentage is almost doubled (17.47\%) among male participants. Age also impacts technology use. Overall, the percentage of older adults who reported not using any \revision{form of} technology in co-located interactions is significantly higher than \revision{with} younger respondents. Specifically, while less than one in ten (8.99\%) Gen Z participants reported not involving technology in their in-person activities, almost one in three (32.14\%) of the Silent participants and 29.48\% of Boomers \revision{indicated that they did not involve} technology in their interactions. One 83-year-old interviewee mentioned that technology is \textit{``not a big part of my life at this point''}:

\begin{quote}
     \textit{``With my grandkids and my daughters and their husbands, I think technology plays a bigger part in their lives than it does in my wife's and my life. Right now technology, uh, is a little confusing to me. At my age. I like it, but I'm not adept \revision{[at]} using it.''} --P15 
\end{quote}

Conversely, technology is an integral part of co-located interactions among younger participants: seven of out ten Gen Z and Millennial \revision{participants} reported using smartphones while hanging out with others (70.22\% and 68.06\%, respectively). We observe that the high percentage of smartphone use among younger population is not always directly linked to the co-located interaction itself. A 20-year-old participant mentioned becoming \textit{``super antsy and nervous''} and feeling that she was \textit{``out of the loop''} if she didn't check her phone periodically:

\begin{quote}
     \textit{``Just because we've grown up with technology, and it's so integrated in our life and everything that we do. It's really difficult for people my age, to set the phone down and kind of be just like in the moment--- and in the moment only--- and not worry about the phone.''} --P4
\end{quote}

\subsubsection{Smartphones are the dominant technology albeit not always the prefer\revision{red} one}
Our survey results show that smartphones are the most used technology in all \revision{of the} top in-person activities, while their use varies depending on the activity people engage in. For example, for those who most enjoy \textit{petting or playing with animals}, about three in four survey participants (73.53\%) reported using a smartphone to \textit{``take pictures,''} while about one in five (20.59\%) mentioned involv\revision{ing} tablets, which were the second most used devices \revision{for participants} hanging out with pets. \textit{Chatting or having conversations} is another activity where smartphones are used more than any other device. Almost two in three (64.93\%) of our survey participants reported using smartphones to \textit{``show each other photos or memes,''} occasionally \textit{``look up things''}, or just be on their phones while casually conversing; a smaller proportion of participants mentioned using a TV (37.44\%) or a computer (21.33\%) while \textit{chatting}. In contrast, among those who most enjoy \textit{intimate or romantic activities}, the \revision{number} of people who reported using smartphones (to \textit{``play music,''} \textit{``watch videos,''} and \textit{``take videos and photos''}) is similar to those \revision{who} cited not using technologies at all (41.12\% versus 38.32\%).

In some cases, however, participants prefer more specialized technologies if they offer better experiences for a group. For instance, P6 mentioned that whenever he has visitors, he always plays music through a smart speaker (e.g., Alexa) to \textit{``set the mood''} and \revision{that he} streams movies from Amazon or Netflix \textit{``instead of passing the phone around.''} Similarly, P22 shared that, while they generally favor a computer when playing video games, they would rather use a gaming console (e.g., an Xbox) when friends come \revision{over} to play together; \textit{``computers can't really have multiple people playing on them, but Xbox has that capability that you can have two people playing at the same time,''} P22 explained.

\subsubsection{Technology is consciously prohibited in some cases and unintentionally \revision{sidelined} in other scenarios}
Among participants who reported not using any technology while hanging out with others, a few stated that technology disappeared without notice because they were \textit{``in the moment.''}

\begin{quote}
     \textit{``We were so in the moment that we didn't really, like, take a picture or take a video... until it's, like, the next day and we were, like, hey, did you take any pictures?''} --P16 
\end{quote}

In other cases, participants intentionally \revision{curtailed} technology use. For example, P25 told us that they have a \textit{``no phone policy''} whenever dining with family and friends because\revision{while it's always possible to,} reach people on the phone, \textit{``you won't always have the opportunity to interact with the people that you're sitting with.''} In some cases, technology can hinder \revision{hinder potential interactions from occurring}: \textit{``When we are out to dinner, he just had his little earpieces in... and whatever he is listening to or texting to or talking to''} (P12). In fact, many noted that when digital devices suddenly appeared \revision{on} the scene, they felt that the individual they were hanging out with \textit{``wasn't really listening''} (P25) or that those moments become \textit{``less genuine.''} (P17). Among them, P6 was exasperated by how frequently his social interactions were disrupted because something appeared on his friend's \textit{``damn phone.''}

\begin{quote}
     \textit{``Honest to God, we would all die if we didn't use our iPhones... even when you're interacting with people, if you're walking around, if you're having a picnic, I'll be darned if we all don't pull out our phones for some stupid reason to either answer a text or look up something... it should be permanently attached to your hand.''} 
\end{quote} 

Interestingly, our findings suggest humans are not the only ones who get frustrated when the other party \revision{becomes} distracted by technology---pets \revision{apparently experience this} too. For instance, P11 described how his two cats \textit{``clearly don't like technology,''} and that they get upset whenever he is on the phone.

\begin{quote}
     \textit{``You can tell by the facial expression; they... look upset and... you can tell by their ears if something's going off. And their tail, you can tell the tail is not, like, up anymore. So you can tell... they want your full attention.''}
\end{quote} 

\subsubsection{Technology is preferable when it affords, sustains, or enriches the activity}
Participants mentioned several different roles that technology plays in co-located interactions. For some, \revision{this has to do with} creating a personal space while enjoying other people's company: \textit{``We would literally sit next to each other... we're both working, headphones on, laptops in our hands... we'll all be co-working in this living room with the couch, TV in the front, and headphones on.''} (P3).
For others, technology helps cultivate the ambiance and set the mood. For instance, P6 mentioned using a smart speaker and a TV when he had visitors, \textit{``you set the stage and you have music. You use those gadgets... Netflix and the smart televisions to watch movies together.''} The \revision{utility} of technology is most evident in activities that involve smartphones. Specifically, the various occasions our interviewees found smartphones useful included \textit{``putting on some music''} to sing and dance with their kids at home (P10), do\revision{ing} a \textit{``mini photoshoot''} to document a memorable trip (P13), finding \textit{``relevant information''} (e.g., news or recipes) to continue a conversation or support an activity (P2), surreptitiously \textit{``chitchatting with friends''} during class \revision{via} text (P26), \textit{``showing each other pictures''} (P18) or \textit{``looking at memes''} to share (P20), exchanging new content encountered on social media to \textit{``start a conversation''} (P4), sending Snaps \revision{during} large gatherings when they \textit{``only want certain people to know this message''} (P4), or filling the gap between activities to avoid boredom:

\begin{quote}
     \textit{``If we sort of run out of things to talk about we'll just put on a movie and sort of, like, chill out... we'll both end up being on our phones a little bit more around each other if we don't really have anything that we need to talk about at that moment.''} --P17 
\end{quote}

\subsection{Value: in-person activities afford physicality and richer nonverbal interactions}
Finally, we explored through the interviews the reasons people \revision{sometimes} prefer in-person activities \revision{to} remote interactions. While many agreed that remote interaction is \revision{\textit{``convenient''}} (P27), \textit{``takes away physical isolation''} (P26), and helps people \textit{``stay updated''} (P10) and \textit{``stay connected''} (P12), participants also reported that they \textit{``get really tired''} (P1), \revision{found} it \textit{``hard to keep up''} (P24), or are more \textit{``likely to get misconstrued''} (P24) when interacting with others online. Participants quoted both pragmatic and emotional dimensions for favoring in-person encounters, while P8 encapsulated the differences between remote and in-person interactions vividly: \textit{``going out and doing things is the relationship... talking to somebody on the phone is upkeep.''}

\subsubsection{Enabling nonverbal cues} 
We see that nonverbal cues, including \textit{``tone of voice''} (P4), \textit{``subtle reactions''} (P17), \textit{``body language''} (P24), and \textit{``facial expressions''} (P26), make in-person encounters preferable \revision{to} interacting through texts, voice calls, or video chats. For example, P4 mentioned that \textit{``there have been many times when we said something and we meant it one way, but the person took it another way because it was a text message... they couldn't read the tone...''} As a result, \textit{``I feel whenever we do have these conflicts, it needs to be talked about in person, just so that nothing is misconstrued.''}

\subsubsection{Facilitating spontaneity}
Many interviewees mentioned that they draw from other ongoing activities or the surrounding environment while engaging in conversations with co-located others; in other words, people tend to talk about \textit{``a lot of the things''} (P24), and the topics are \textit{``spontaneous''} (P1). Spontaneous conversations that people enjoy, however, are less likely to happen through online platforms. For example, P1 noted that \textit{``digital tools... kind of contain or create a boundary around the conversation,''} which may pose challenges \revision{to} connecting \revision{with} people, to the extent that P27 shared how much they disliked talking to a friend through text messaging: 

\begin{quote}
     \textit{``He's not really that great of a texter. He'll text, like, short sentences and stuff... a better way to describe it is just the way that he texts, it doesn't feel like I'm talking to [him]... if that makes sense. Like, I feel like I'm talking to a different person''}
\end{quote}

\subsubsection{Supporting authenticity}
Many interviewees noted that they prefer in-person interactions because they are \textit{``more raw''} (P21). For example, P23 noted that \textit{``just having a screen between you... takes away... the essence of a person.''} Similarly, P25 agreed that the \textit{``true emotions or true reactions''} often get lost on digital platforms. Overall, we observed that on-the-spot nonverbal reactions, given their immediacy and unmediated nature, play a key role in making in-person interactions more genuine. According to P17,

\begin{quote}
     \textit{``Just the fact that you have less time to process and think through... makes a big difference... there's, like, a different weight placed on those words when you're texting someone than if you say it in-person.''}
\end{quote}

\subsubsection{Encouraging undivided attention}
P22 noted that, \revision{in contrast} to talking over voice chat and not knowing if others are paying attention, he preferred \textit{``having the confirmation''} that people are \textit{``actually there... listening and responding''} so that he felt \textit{``safer''}. \revision{Many also} noted that having others' \textit{``undivided attention''} makes them prefer interacting in-person: \textit{``when you're FaceTiming with someone, not all of your attention is just on them because there are other things going on around you. But when you're with someone, it's kind of hard not to just sit down and give them your undivided attention''} (P16). Additionally, P1 saw the chance to be fully engaged with others in co-located scenarios \revision{as} central to relationship building: \textit{``the fact that [our] being physically co-present together is the thing that keeps our relationship going. I guess because if we weren't, it would be easy to drift apart.''}

\subsubsection{Leveraging physicality}
Finally, participants mentioned that being able to engage in physical interactions makes in-person encounters more fun, intimate, and enjoyable. For example, P3 described that \textit{``I'm a very physical person. You know, I'm not a really good texter, I'm not a good messager... I like the physicality of being with people.''} Similarly, P25 mentioned \textit{``I'm very, like, huggy and cuddly person''} and \revision{that she} enjoyed physical interactions. She continued, 

\begin{quote}
     \textit{``Me and my friend, sometimes we'll wrestle and, like, physically interrupt each other in the form of wrestling, chasing each other, or just like that... on FaceTime, all we can really do is just talk.''}
\end{quote} 

To others, simply being physically co-located helps support the sense of connectedness. As P24 put it, \textit{``It's just like sitting next to somebody, you're, like, still connected, even if you're not talking... so you can, like, foster conversation after just, like, being there.''} Likewise, P27 saw in-person interactions \revision{as being} more \textit{``heartfelt,''} noting that \textit{``I feel like I connect better with my friends when I see them in-person.''}

\section{Discussion}
Our results show that different contextual factors---including the \textit{relationship} between individuals who hang out together, the \textit{activity} they engage in, the \textit{location} \revision{in which} social encounters take place, and the \textit{technology} involved---all create value that make co-located social interactions enjoyable. We summarize our results and \revision{present our} design recommendations for co-located technology design in Table~\ref{tab:results-summary}.

\begin{table}[th]
\renewcommand{\arraystretch}{1.3}
\caption{Summary of our study findings and design recommendations for designing enjoyable co-located technologies}
\label{tab:results-summary}
\centering
\resizebox{\textwidth}{!}{%
\begin{tabular}{{p{0.13\textwidth}p{0.43\textwidth}p{0.43\textwidth}}}
\toprule
\textbf{Context} & \textbf{Study Findings} & \textbf{Recommendations} \\
\midrule
Relationship & Participants reported that they most enjoy spending time in-person with their \textbf{strong social ties} (e.g., romantic partners, friends, children, and pets).
& Focus design efforts on supporting relationships between \textbf{``besties''} by maintaining and strengthening \revision{persistent} social connections. \\ 
\hline
Activity & Participants most enjoy having \textbf{day-to-day, light-weight} social interactions. Multiple in-person activities often happen \revision{simultaneously.} 
& Leverage and \textbf{``piggyback''} existing everyday activities rather than creat\revision{ing} new ones. Design social artifacts that allow multitasking. \\

\hline
Location & The most enjoyed in-person activities either took place in \textbf{intimate private spaces} or were \textbf{not tied to a location at all}. & Design co-located technologies to complement a private space's \textbf{ambiance} or design them to be portable \textbf{(``in-a-box'')} and usable anywhere. \\

\hline
Technology & People often used digital devices to \textbf{initiate} social encounters but do not want technologies to disrupt or supersede ongoing face-to-face interactions. & Design digital technologies that work as \textbf{``board games'':} systems that catalyze social interactions by attracting a shared locus of attention. \\

\bottomrule
\end{tabular}
}
\end{table}

\subsection{Relationship: designing for long-term relationships between besties}

While we cannot design a person \revision{that} people enjoy hanging out with, as HCI researchers, we can design systems to make in-person interactions engaging and enjoyable. Unsurprisingly, our results show that people much prefer spending time with their ``strong ties'' (e.g., friends) than ``weak ties'' (e.g., acquaintances) \cite{granovetter1977strength}.
This finding aligns \revision{well with} Olsson et al.'s recent work \revision{highlighting} a research gap in \revision{studies focused on} designing co-located technologies to ``sustain interaction that has already initiated'' (rather than focusing efforts on ``initiating encounters between strangers'') \cite{olsson2019technologies}. Taken together, we see untapped potential in designing technologies that maintain and enhance long-term relationships that may otherwise fade away or banal over time. We outline two specific recommendations below.

\subsubsection{Encouraging group-specific in-person interactions}
As P3 puts it, \textit{``we didn’t just build memories, we built routines.''} Our study suggests that intimacy is often cultivated through the establishment of routine-like activities that are distinctive, while participants hang out with different groups of people. As a result, we suggest future work focus on supporting group-specific encounters \revision{that} maintain and support intimacy between besties. For instance, for couples who enjoy cooking together, we can build interactive couples cooking apps that \revision{provide them with} different cooking instructions so that the couple can work together to put together a perfect meal. Alternatively, (inspired by the classic children’s game hide-and-seek and the mobile running app Zombies, Run! \cite{Zombierun}) for friends who enjoy running outdoors together, we can build mobile systems that give audio cues where randomly assigned “seekers” have to locate and catch concealed “hiders” within a given time to win the game.

\subsubsection{Designing multi-species gaming experiences}
One aspect we find surprising is that pets are \revision{among} the top\revision{-ranked} individuals people enjoy spending in-person time with, especially when playing games (10.94\%). Prior works such as Cat Cat Revolution \cite{noz2011cat} and BubbleTalk \cite{ko2018bubbletalk} \revision{represent} systems that incorporate multimedia simulations to facilitate collective play between people and their pets (i.e., cats and fishes). Research on animal-computer interaction has gained traction in HCI \cite{mancini2011animal}, but there are few systems that aim to support human-animal gaming experiences. One potential area for co-located technologies is incorporating mobile AR to make traditional games more challenging and fun. For example, inspired by the classic video game Super Mario \cite{Supermario}, we can design a mobile treasure treat game where a dog collects AR coins that appear to be scattered on the ground \revision{while} playing fetch or an AR agility competition game where the dogs and their owners pair up to perform certain tricks and compete against other teams.

\subsection{Activity: leveraging existing, light-weight social interactions}
We observe that many in-person activities our participants enjoy are mundane casual interactions; these activities may seem trivial but are, in fact, fundamental to relationship building. Additionally, our findings suggest that people are drawn to mundane activities, not only because of the emotional fulfillment they offer, but also because everyday activities are easily accessible. Similarly, a recent article \revision{in the} New York Times also suggests that \revision{in} ``life for anyone but the very rich---the physical experience of learning, living and dying---is increasingly mediated by screens'' \cite{humancontact}. Combining these insights, we recommend that future work focus on leveraging and augmenting activities that are already happening---a technique known as ``piggyback prototyping'' \cite{grevet2015piggyback}.

One potential opportunity for co-located technologies is to make routine activities more collaborative and engaging; as such, future work can benefit from creating group challenges to gamify house\revision{hold} chores. For example, inspired by the maze arcade game Pac-Man \cite{pacman}, we can build an AR floor-mopping challenge where parents and their children collect virtual coins and avoid \revision{coming} in contact with any AR ghosts that move in the space while \revision{they are} mopping the floors together (building on the finding that 32.8\% of our participants prefer playing with their children).

\subsection{Location: designing for ambiance or in-a-box experiences}

Our results show that many popular in-person activities such as \textit{chatting} and \textit{eating} can take place anywhere; in cases where an interaction is tied to a specific location, it most likely takes place in private spaces. The HCI community has long differentiated the concepts of ``space'' and ``place'': ``space'' focuses on the factual geometrical arrangements\revision{, while} ``place'' emphasizes the constructed and perceived meaning of an environment \cite{harrison1996re, dourish2006re}. Building on the emotional connection inscribed in the notion of a place, we see an opportunity to design co-located technologies that focus on placemaking \cite{freeman2019smart, chamberlain2016audio}; in different contexts, prior research has suggested strategies such as garnering emotional connections to strengthen place attachment \cite{milligan1998interactional} and encouraging multisensory discovery of the environment \cite{chamberlain2016audio, lentini2010space}. We outline two potential directions that aim at supporting these two distinctive location preferences in this section.

\subsubsection{Complementing the ambiance of private spaces}
We notice that participants preferred private space when engaging in activities that are more intimate (e.g., \textit{intimate or romantic activities} and \textit{resting}) or require particular setups (e.g., gaming consoles and big screens for \textit{playing games}, kitchen equipment for \textit{cooking or baking}, TV and audio system for \textit{watching, listening, and reading}). More specifically, participants described the private settings they prefer as \textit{``a laid-back, quiet environment''} (P13) where they can sit \textit{``next to each other''} (P14), being \textit{``close to music''} (P3), while still \revision{being} \textit{``able to actually hear each other''} (P4). Accordingly, we suggest that designers pay close attention to the atmosphere and setting specific private spaces afford to create co-located technologies that support location-dependent activities.

\subsubsection{Designing in-a-box experiences for location-independent scenarios} 
Our results show that many \revision{of the} top in-person activities such as \textit{chatting}, \textit{eating}, \textit{traveling}, and \textit{partying} are location-independent. From this, we see an underexplored opportunity to build co-located technologies that mobilize activities that are often tied to a specific location. One specific strategy for transforming location-dependent activities into interactions that can happen on the move or are location-independent is to bring these existing experiences in-a-box. One of the few designs that tapped into this space included Roxie \cite{Roxie}, a plug-and-play karaoke station \revision{that enables} people to sing together \revision{while} carpooling (as opposed to going to a Karaoke bar). Another example is Escape Room In A Box \cite{escaperoom}, a board game kit that allows people to play a room escape game anywhere (instead of having to visit an actual escape room). Future designs \revision{may} aim at packaging location-specific experiences in a box, \revision{allowing} them to be played at anywhere.

\subsection{Technology: designing digital devices as ``board games''}
Digital devices have become an indispensable part of our \revision{everyday lives}. However, our study shows that establishing a consensus among co-located individuals regarding \textit{how} and \textit{when} to involve technology is critical in order to avoid frustration and discomfort. Specifically, \revision{we observe} that participants considered technology undesirable when it \textit{``supersede(s) a conversation''} (P6) or disrupts ongoing interactions. Conversely, technology is preferable when it helps initiate, maintain, or enrich in-person activities, whether \revision{this} is about \textit{``setting the stage''} through music (P6), \textit{``filling the gap''} between interactions (P17), or capturing the moment by \textit{``taking pictures and videos''}(P11). \revision{With this in mind,} instead of creating devices that distract people from their collective experiences, we suggest building co-located technologies that function more \revision{like} board games, which are at the center \revision{of the activity} drawing everyone's attention to one another. 

A few previous \revision{studies have presented} theoretical concepts and design strategies that align with our recommendations. For example, to address the issue of phubbing, Isbister coined the theoretical concept of ``suprahuman'' to focus on the spaces between co-present individuals \revision{in order to create} technologies that ``help to build webs of connection and coordination and co-action'' \cite{isbister2019toward}. Similarly, Oduor et al. suggest making visible the contents on a personal device to co-located family members when situations that require immediate attention occur to avoid a sense of disengagement \cite{oduor2016frustrations}.Building on this line of work, we see untapped potential \revision{for} creating designs that resemble existing technologies that already bring people together (e.g., TV, gaming console, or audio systems). The public installation Giant Sing Along \cite{giant-sing-along} is an interesting example that combines an oversized screen that displays song lyrics (resembling the function of a TV) and an array of standing microphones (resembling the function of an audio system), invit\revision{ing} co-located individuals to sing together and to create a collective musical experience. Overall, considering the fact that technology is already deeply embedded in our day-to-day interactions, instead of curbing the use of digital devices, we believe that future work should bring technology to the center to emphasize and enhance the collective experience.

\subsection{Value: leveraging physicality to emphasize the sense of ``being there''}
Our results suggest that remote interactions may better function as maintaining the social ties between individuals. In contrast, people prefer in-person activities because they offer opportunities to draw from nonverbal communication, engage in spontaneous interactions, express authenticity and attentiveness, and participate in physical interactions. Prior work on social presence aims to promote ``the sense of being with another'' through technology \cite{biocca2003toward, ijsselsteijn2003staying} and outlines design concepts, such as embodiment and immediacy \cite{biocca2002defining, van2016engagement}. Our results align with previous research \revision{and suggest that} \textit{physicality} is the central element in facilitating the sense of connectedness and intimacy \revision{of in-person interactions}. More specific examples that foreground the notion of physicality as an enjoyable element of in-person interactions include descriptions such as being \textit{``next to each other''} (P3), feeding \textit{``each other's energy''} (P13), and having \textit{``physical touch''} (P24) while hanging out with those close to us.

We suggest designing technologies that leverage physicality---such as physical proximity, bodily movements, vocal tones, and facial expressions---to enhance the sense of ``being together.'' One technique that has been previously explored is track\revision{ing} and amplify\revision{ing} the biosignals (e.g., heart rate) among those who are co-located to support intimacy \cite{howell2019life}, cultivate connectedness \cite{min2014biosignal}, or augment collective play \cite{frey2016remote}. Application-wise, \revision{in discovering} that many of our participants regret forgetting to take photos or videos to capture their in-person interactions (see \S4.4.3)\revision{, we have considered the potential for also incorporating sensors for monitoring} biosignals in image/video capturing devices to trigger their recording functions. Such a mechanism, outlined by Metz \cite{Metz2018} in a different context, \revision{could} help document co-located experiences without preventing people from being fully engaged in their interactions. \revision{In addition to} biosignals, designers can also leverage other dimensions of physicality to enhance co-located experiences. For instance, inspired by the mobile laser tag-like game QuasAR Arena \cite{quasar}, we can design co-located technologies that specifically aim to support participants who enjoy chasing one another (see \S4.5.5) by building AR snow fight games that prompt multiple players to cover their competitors with virtual snowballs while running and hiding to dodge snowballs themselves.

\section{Limitations}
Several aspects of our study design \revision{may} have influenced our results. First, while our survey sample aimed to reflect the demographics of the United States, the results may not generalize to populations in other countries. In addition, our interviewees, although diverse in gender, age, race, and geographical location, \revision{were} slightly skewed \revision{towards} younger participants (40.74\% of our interviewees were Gen Z participants). We suggest following a similar interview protocol to gather more in-depth insights prior to designing co-located technologies for older adults. Further, we note that we did not collect \revision{the} relationship status of our participants in this study. While not having access to this piece of information does not affect or change our results, we acknowledge that knowing whether a participant is currently in a relationship can yield additional insights (e.g., more Gen Zs prefer spending time with friends because many of them do not have romantic partners). \revision{Additionally, it is worth noting that this paper maintains a critical distance with regards to technology. We do not focus solely on technology’s impact on social interactions, but we do dive into its role when participants mention it, as it is the case in §4.4.3.}

Another limitation of this study relates \revision{to the way} COVID-19 \revision{may} have restricted in-person activities. Specifically, when we started our interviews, a few states (e.g., California and New York) \revision{had} just begun to issue stay-at-home orders due to the COVID-19 outbreak. Since our goal was not comparing people's preference for social interactions before and after the pandemic, during the interview onboarding, we explicitly instructed our participants to share the in-person interactions they enjoyed when social gatherings were not constrained by the pandemic. Further, we also asked our participants to describe in detail \textit{``the last time''} they engaged in their preferred co-located interactions to help minimize biases in \revision{our} data (see the full interview script in the Appendix). Following this approach, we note that a different study is needed to design co-located technologies for a post-pandemic ``new normal'' \cite{Lichfieldarchive2020}. Despite the limitations, we believe that our work---including our study design and research findings---provides important groundwork to guide future user studies and technological development in supporting enjoyable in-person social interactions through technology.

\section{Conclusion}
Through a mixed-method study, this work explored how different contextual factors---relationship, activity, location, and technology---play a part in making in-person interactions enjoyable. We observed that people prefer having everyday casual social interactions with their strong ties, and the relationship between co-located individuals \revision{is} a deciding factor \revision{in} the activity, location, and technology involved in in-person activities. In addition, our results also show that physicality (e.g., nonverbal cues, bodily interactions), spontaneity, and authenticity not only supports connectedness but also helps co-located individuals be in the moment.

This work provides an empirical foundation for a new and under-explored design space: creating technologies that support enjoyable co-located interactions.

\received{January 2021}
\received[revised]{July 2021}
\received[accepted]{November 2021}

\begin{acks}
\revision{We thank our colleagues at Snap Research, especially Maarten Bos, Fannie Liu, and Ella Dagan for their thoughtful comments in our initial manuscript. Thanks also to the ACs and reviewers for their excellent feedback, which greatly improved the earlier version of this paper. This work is supported in part by Snap Inc. and the U.S. National Science Foundation (Grant \#2030856, the Computing Research Association for the CIFellow Project\footnote{Disclaimer: Any opinions, findings, and conclusions or recommendations in the material are those of the author(s) and do not necessarily reflect the views of the National Science Foundation nor the Computing Research Association.}) that supported Liu in her postdoctoral position, when she worked on this paper after working at Snap.}
\end{acks}

\appendix
\section*{Appendix}
\section{Survey Questions}
\label{sec:survey}

The following is a complete list of the questions we asked in our survey.

\begin{enumerate}
\item With whom do you most enjoy spending time \textbf{in person}? (Single choice,
randomize choices)
\begin{itemize}
	\item Romantic Partners
	\item Parents
	\item Siblings
	\item Children
	\item Friends
	\item Acquaintances
	\item Schoolmates
	\item Co-workers
	\item Neighbors
	\item Strangers
	\item Other (please specify):
\end{itemize}

\item When spending time with \textbf{your [insert answer from question 1] in person}, what do you enjoy doing together? Select up to 5 that apply. (Multiple choices, randomize choices)
\begin{itemize}
	\item Chatting or having conversations
	\item Cooking or baking
	\item Doing chores
	\item Drinking or smoking
	\item Eating
	\item Exercising, doing sports, or dancing
	\item Going to cultural events
	\item Goofing around or pranking others
	\item Having intimate or romantic activities
	\item Making arts and crafts
	\item Making videos or taking photos
	\item Messing around on our phones
	\item Partying or socializing
	\item Petting or playing with animals
	\item Playing games
	\item Practicing spirituality
	\item Resting
	\item Shopping
	\item Teaching and learning
	\item Traveling or sightseeing
	\item Video chatting with others
	\item Volunteering
	\item Wandering around
	\item Watching, listening, or reading
	\item Working
	\item Other (please specify):
\end{itemize}

\item Please rank the activities you enjoy doing with your \textbf{[insert answer from question 1] in person} in order of preference (drag and drop to have your favorite activities at the top).

\item Are you usually at a specific location while \textbf{[insert favorite activity from question 3]} with your \textbf{[insert answer from question 1]} in person?
\begin{itemize}
	\item Yes, we only do it at a specific location
	\item No, we do it anywhere (Skip next question)
\end{itemize}

\item How would you describe the type of location while \textbf{[insert favorite activity from question 3]} with your \textbf{[insert answer from question 1]} in person?
\begin{itemize}
	\item Public (e.g., mall, park, library)
	\item Private (e.g., home, private vehicle)
	\item Semi-public (e.g., office, community courtyard)
	\item Other (please specify):
\end{itemize}

\item Some people report using technology in some form when they are together. Reflecting back on your favorite activity, \textbf{[insert favorite activity from question 3]}, with your \textbf{[insert answer from question 1]}, which of the following devices are also involved in the activity? Please select all that apply.
\begin{itemize}
	\item Smartphone
	\item Smart watch (e.g., Apple Watch, Fitbit)
	\item Smart speaker (e.g., Amazon Echo, Google Home)
	\item Audio system or speaker
	\item TV
	\item Computer
	\item Tablet (e.g., iPad, Fire)
	\item Headphones or earbuds (e.g., Apple AirPods, Bose SoundWear)
	\item Gaming console (e.g., Xbox, Nintendo Switch, PS4)
	\item VR/AR headset (e.g., Oculus Rift, HTC Vive, Magic Leap)
	\item None of the above
	\item Other (please specify):
\end{itemize}

\item Could you tell us more about how the last time you and your \textbf{[insert answer from question 1]} did your favorite activity, \textbf{[insert favorite activity from question 3]}? What exactly did you do? Where did you do it? What tools, devices, or other stuff, if any, did you use? Be as detailed as you can. (Free text with a minimum of 30 characters)

\item What is your age? 

\item Which gender do you identify with? (Multiple choices)
\begin{itemize}
	\item Female
	\item Male
	\item Non-binary
	\item Prefer not to answer
	\item Prefer to self-describe: 
\end{itemize}

\item What is your race or ethnicity? 
\begin{itemize}
	\item American Indian or Alaska Native
	\item Asian
	\item Black or African American
	\item Hispanic or Latino
	\item Native Hawaiian or Other Pacific Islander
	\item White
	\item Two or more races
	\item Prefer not to answer
	\item Other (please specify):
\end{itemize}

\item What is your primary occupation? 
\begin{itemize}
	\item Administrative Support (e.g., secretary, assistant)
	\item Art, Writing, or Journalism (e.g., author, reporter, sculptor)
	\item Business, Management, or Financial (e.g.,manager, accountant, banker)
	\item Computer Engineering or IT Professional(e.g., programmer, IT consultant)
	\item Education or Science (e.g., teacher, professor, scientist)
	\item Engineer in other field (e.g., civil or bioengineer)
	\item Homemaker
	\item Legal (e.g., lawyer, paralegal)
	\item Medical (e.g., doctor, nurse, dentist)
	\item Qualtrics worker
	\item Self-employed
	\item Skilled Labor (e.g., electrician, plumber, carpenter)
	\item Retired
	\item Unemployed
	\item Middle or high school student
	\item College student
	\item Graduate student
	\item Prefer not to answer
	\item Other (please specify):
\end{itemize}

\item Which of the following best describes your highest achieved education level? 
\begin{itemize}
	\item Some High School
	\item High School Graduate
	\item Some college (no degree)
	\item Associates degree
	\item Bachelors degree
	\item Graduate degree (e.g., Masters, Doctorate)
	\item Prefer not to answer
	\item Other (please specify):
\end{itemize}

\item We are part of a research team looking to understand more about where, why, how and with whom people hang out in-person. Are you interested in participating in an online video interview (via Google Hangout) to tell us more about time you spend in-person with others? The interview is expected to take about 60 minutes and you will receive a \$20 Amazon gift card after completing the interview. We will require your name and email to sign up and will contact you via email if you are selected to participate in the online interview.
\begin{itemize}
	\item Yes
	\item No (skip to the last question)
\end{itemize}

\item What is your preferred email address for us to contact you? Your email address will only be used to contact you about the following interview opportunities:

\item Comments, questions, or notes for us: 

\end{enumerate}

\section{Interview Protocol}
\label{sec:interview-protocol}

The following is the script--- including the on-boarding introduction and start-off questions--- we used as during our semi-structured interviews.

\subsection{On-Boarding Introduction}
Hi, [insert participant's name], thank you for agreeing to participate in our interview. I’m [first author's name], a researcher from [institution], and we’re interested in understanding what people enjoy doing while they’re spending time with others in-person. This will help us create technologies in the future to support in-person activities. 

I know that things might have changed quite a bit in recent weeks, and I’m happy to hear about it if you’re willing to share, but I’m most interested in learning about how you spend time with others in a typical day-to-day scenario. There are no right or wrong answers to any of the questions I will be asking, I just want to know more about what you enjoy doing while hanging out with others in person.

If there is any point during the interview that you have a question, want to take a break, or wish to drop out, please let me know, it will be totally fine. During the interview, I will take notes, as well as record the audio and video of our conversation. You have previously signed and returned the participant consent form, but before we begin, I’d like to ask again if you consent to be recorded for this interview? [Wait until the participant consent to be recorded]. Ok, I will start recording now.

\subsection{Warming Up and Recap}
Before we start, can you tell me a little bit about yourself? Feel free to tell me anything you feel comfortable sharing. 
\begin{itemize}
	\item How old are you?
	\item What do you do?
	\item What do you like to do for fun?
	\item How would you describe the place you live?
\end{itemize}

In the survey, I think you mentioned that you enjoy spending time in-person with [insert answer from survey question 1], can you tell me a little bit more about it?
\begin{itemize}
	\item What do you enjoy doing with them?
	\item How did you guys meet?
	\item How would you describe your relationship with them?
\end{itemize}

\subsection{Follow-Up Questions}
\subsubsection{Location}
Tell me more about the last time you did [activity] with [person]. 
\begin{itemize}
	\item Where were you?
	\item Tell me more about this specific place. How do the settings look like and what are some objects around you? 
	\item Why did you two pick that as the place to do [activity]?
	\item In what other places have you done [activity] together? 
	\item How would you imagine doing [activity] in a different place?
\end{itemize}

\subsubsection{Time}
Can you walk me through the process when you last did [activity] with your [person]? 
\begin{itemize}
	\item How did you plan it?
	\item What did you do while [activity]?
	\item What happened after [activity]?  
\end{itemize}

\subsubsection{Technology and Artifacts}
I'd like to know more about the things you need to do [activity]. 
\begin{itemize}
	\item What are some objects, devices, or tools you use while doing [activity]?
	\item Do you use any technology while doing [activity] with [person]?
	\item Imagine doing [activity] with [person] while not having [technology or artifact], how would you imagine it to be?
	\item This question might be a little abstract, but how would you describe the role of the [technology or artifact] while you two [activity] together?
	\item You have mentioned a few objects that help you and your [person] do [activity]. Can you tell me about the objects or things that get in the way of that activity? 
\end{itemize}

\subsubsection{Relationship and Activity}
You mentioned that you enjoy doing [activities] with [person].
\begin{itemize}
	\item Do you do the same activities with other people? How does it feel similar/different compared to doing it with the person you most enjoy spending time in person with?
	\item Besides [activities], what else do you enjoy doing with [person]?
	\item Tell me more about the type, quality, and the characteristics of the [activity].
	\item Describe a memorable or unique time when you do [activities] with [person]. What makes it special? Is there any specific moment like this that stands out?
	\item Can you describe a time when [activities] with [person] were not as enjoyable? What happened?
\end{itemize}

\subsubsection{In-Person Interactions} Tell me about how you interact with [person] when you are not physically together.
\begin{itemize}
    \item What differences do you see between hanging out with [person] in-person and online?
	\item Do you prefer hanging out with [person] in-person or online? Why?
\end{itemize}
 

\subsection{Closing}
That’s about it for today. Do you have any questions for me? 
Thank you for participating in the interview today, you will receive the gift card from us shortly after. If you do not receive it in the next three days, please email me to let me know. Enjoy the rest of your day.

\bibliographystyle{ACM-Reference-Format}
\bibliography{CSCW21}

\end{document}